\documentclass[aps,prd,twocolumn,a4paper,superscriptaddress]{revtex4-1}
\usepackage{graphicx}
\usepackage{multirow}
\usepackage{latexsym}
\usepackage{hyperref}
\usepackage[british]{babel}
\begin{document}

\newcommand{\be}{\begin{equation}}
\newcommand{\ee}{\end{equation}}
\newcommand{\bq}{\begin{eqnarray}}
\newcommand{\eq}{\end{eqnarray}}
\newcommand{\bsq}{\begin{subequations}}
\newcommand{\esq}{\end{subequations}}
\newcommand{\bc}{\begin{center}}
\newcommand{\ec}{\end{center}}

\title{Cosmic strings and other topological defects in nonscaling regimes}

\author{R. P. L. Azevedo}
\email[]{up201109025@fc.up.pt}
\affiliation{Faculdade de Ci\^encias, Universidade do Porto, Rua do Campo Alegre 687, 4169-007 Porto, Portugal}
\author{C. J. A. P. Martins}
\email{Carlos.Martins@astro.up.pt}
\affiliation{Centro de Astrof\'{\i}sica da Universidade do Porto, Rua das Estrelas, 4150-762 Porto, Portugal}
\affiliation{Instituto de Astrof\'{\i}sica e Ci\^encias do Espa\c co, CAUP, Rua das Estrelas, 4150-762 Porto, Portugal}
\date{30 January 2017}

\begin{abstract}
Cosmic strings are topological defects possibly formed in the early Universe, which may be observable due to their gravitational effects on the cosmic microwave background radiation or gravitational wave experiments. To this effect it is important to quantitatively ascertain the network properties, including their density, velocity or the number of strings present, at the various epochs in the observable Universe. Attempts to estimate these numbers often rely on simplistic approximations for the string parameters, such as assuming that the network is scaling. However, in cosmological models containing realistic amounts of radiation, matter and dark energy a string network is never exactly scaling. Here we use the velocity-dependent one-scale model for the evolution of a string network to better quantify how these networks evolve. In particular we obtain new approximate analytic solutions for the behavior of the network during the radiation-to-matter and matter-to-acceleration transitions (assuming, in the latter case, the canonical $\Lambda$ cold dark matter model), and numerically calculate the relevant quantities for a range of possible dark energy models.
\end{abstract}

\maketitle

\section{Introduction}\label{sec:intro}

Topological defects are physical structures produced in symmetry-breaking phase transitions in the early universe first theorized by Kibble \cite{Kibble}. Among all the possible types of defects, the one-dimensional cosmic strings are the focus of most of the studies in this area \cite{Vilenkin}, due to their compatibility with current cosmological models and their association with several brane inflation scenarios \cite{SarangiTye,PolchinskiCopelandMyers} and supersymmetric grand unified theories \cite{Rocher}. Currently they are mostly constrained by the cosmic microwave background \cite{Landriau,Planck,Lazanu,Ringeval,Charnock,Kunz,Hergt}, but in the coming years they will also be constrained by gravitational wave facilities \cite{Damour0,DamourVilenkin,Siemens,Olmez,Binetruy}.

The general features of the cosmological evolution of topological defect networks are well understood, and can be quantitatively described using the velocity-dependent one-scale (VOS) model \cite{MartinsShellard,MartinsShellard2,VOSbook}, an extension of Kibble's one-scale model \cite{Kibble2}. Broadly speaking, in cosmological scenarios with a single component string networks rapidly become relativistic and evolve in the well-known linear (scale-invariant) scaling regime. The attractor nature of this solution, and the fact that in this case the scaling velocity is a substantial fraction of the speed of light are well established, for example in the radiation and matter eras. For other early approaches to analytic modeling, see Refs. \cite{Kinks,Vachaspati,Perivolaropoulos,Austin}.

However, realistic universes are not as simple as that, containing transitions between radiation and matter domination, and later between matter and dark energy domination. It is perhaps less understood that around the former a string network is {\it not} scaling, and indeed that in an accelerating universe no scaling solution exists at all. Given that it is common in the literature to make the assumption that the network is scaling throughout its evolution, one may ask how good this approximation is. This is particularly pertinent in light of forthcoming higher-quality data, which need to be matched by accurate theoretical predictions and templates.

Here we address this issue, and use the VOS model to better quantify how these networks evolve. For both the radiation-to-matter and the matter-to-acceleration transitions we obtain new approximate analytic solutions for the behavior of defect networks which interpolate between the well-known asymptotic scaling solutions in each of the eras. These solutions are valid for defects of any dimension: in addition to the obvious conformal time or redshift dependence, they also explicitly depend on the defect dimension and on the phenomenological parameters of the VOS model.

For the end of matter domination and the onset of acceleration, the nature of dark energy is a known unknown. Therefore, while we assume a $\Lambda$ cold dark matter ($\Lambda$CDM) model to obtain the aforementioned analytic solution, we will also numerically solve the VOS equations for the specific case of cosmic strings and explore the dependence of the VOS model variables---the network's correlation length and root mean square (RMS) velocity---on the cosmological parameters characterizing the dark energy model. A further quantity of interest is the number of cosmic strings present in the observable universe \cite{Consiglio}, which is relevant for inferring many of the observable consequences of these networks, such as their possible observation though gravitational lensing.

\section{Velocity-dependent one-scale model}\label{sec:model}

The VOS model provides a quantitative and physically clear methodology to describe the evolution of topological defect networks, and has been calibrated against numerical simulations. Here will simply provide a concise overview, introducing the aspects that will be relevant for our subsequent analysis. For more detailed descriptions we refer the reader to the recent overview \cite{VOSbook}.

\subsection{Canonical VOS model}

The model arises from considering that the network can be described by averaged quantities, such as its energy $E$ and RMS velocity $v$ \cite{MartinsShellard,MartinsShellard2}. The term 'one scale' expresses the assumption that the correlation length and defect curvature radius coincide with a characteristic length scale $L$. Let us consider the generic case of defects with an $n$-dimensional worldsheet and start by defining this characteristic length scale
\be
L^{4-n}=\frac{M}{\rho}\,, \label{charscale}
\ee
where $M$ will have dimensions appropriate for the defect in question (i.e., monopole mass, string mass per unit length, or wall mass per unit area), and can also be written
\be
M\sim \eta^n, \label{massscale}
\ee
with $\eta$ being the corresponding symmetry-breaking scale. For the case of cosmic strings ($n=2$) the relevant parameter will be the string mass per unit length, denoted $\mu$.

For simplicity, in what follows we will ignore the effects of friction due to particle scattering, which typically are only relevant in epochs of cosmological defect evolution that are much earlier than can be probed by astrophysical observations. Under these assumptions one can derive the following evolution equations for the characteristic length scale $L$ and RMS velocity $v$ \cite{VOSbook}
\be
(4-n)\frac{dL}{dt}=(4-n)HL+nHLv^2+cv
\ee
\be
\frac{dv}{dt}=(1-v^2)\left[f-nHv\right]\,.
\ee
There are two dimensionless parameters, $c$ describing the energy losses by the network (in the case of cosmic strings this is the loop chopping efficiency), and the momentum parameter $k$ characterizing the defect curvature. In the velocity equation the generic $f$ term describes driving forces affecting the defect dynamics. For extended objects (walls and strings) that have been studied in detail in the past, this driving force is just the local curvature, and we have
\be
f\sim\frac{k}{R}=\frac{k}{L}\,;
\ee
in the last equality we are implicitly assuming that our characteristic length scale $L$ is the same as the defect curvature radius $R$, consistently with our one-scale assumption.

In a universe in which the scale factor grows as $a\propto t^\lambda$, with $0\le\lambda<1$,  one can then show that the attractor (asymptotic) solution of this system of equations is the linear scaling solution
\be
L=\epsilon t\,,\quad v=v_0=const.
\ee
with the proportionality factors depending on the model parameters and the expansion rate as follows
\be\label{stdL}
\epsilon^2=\frac{k(k+c)}{(4-n)n\lambda(1-\lambda)}\equiv\frac{m^2}{\lambda(1-\lambda)}
\ee
\be\label{stdV}
v_0^2=\frac{4-n}{n}\frac{1-\lambda}{\lambda}\frac{k}{k+c}\,.
\ee
However no such solution exists if the scale factor is not a power law, as is the case in the transition between the radiation and matter eras, or at the onset of the recent acceleration phase. Note that in the former equation we have, for future convenience, defined the parameter
\be\label{defm2}
m^2\equiv\frac{k(k+c)}{(4-n)n}\,,
\ee
which will depend on the dimension of the defect---apart from the explicit dependence on $n$, the phenomenological parameters $c$ and $k$ are expected to depend on it too.

We will also be interested in the number of cosmic strings in the observable universe at a given time (or redshift), $N(z)$, which can be calculated assuming that on average each cube with side $L(z)$ contains one string. Therefore $N(z)$ can be obtained from
\begin{equation}\label{number}
N(z)=8\left[\frac{d_{H}(z)}{L(z)}\right]^3\,,
\end{equation}
where $d_{H}(z)$ is the size of the horizon, given as a function of redshift by
\begin{equation}\label{horizon}
d_{H}(z)=\frac{1}{1+z}\int\frac{dz'}{H(z')}\,.
\end{equation}
This number of string segments $N(z)$, and in particular its present-day value, is needed for accurate estimates of astrophysical signatures of these networks, including its effects as gravitational lenses and in cosmic microwave background maps.

\subsection{Invariant and physical quantities}

Throughout the previous subsection, the characteristic length scale $L$ was an invariant quantity or, in other words, a measure of the invariant string energy (and hence length). We now discuss how to express the VOS model in terms of a physical length scale, which will be useful for some of our subsequent analysis. A more thorough discussion can be found in Ref. \cite{Cabral}. Invariant and physical quantities are related through the standard Lorentz factor, $\gamma=(1-v^2)^{-1/2}$; for energies this is simply
\be
E_{inv}=\gamma E_{ph}\,,
\ee
and since, according to Eq. (\ref{charscale}), the defect density is $\rho\propto L^{-(4-n)}$ we see that the characteristic length scales are related via
\be
L_{ph}=\gamma^{\frac{1}{4-n}}L_{inv}\,. \label{relphinv}
\ee
Note that this length scale is a measure of the total energy content of the network, or (in the context of the VOS model assumption of a single independent characteristic scale) the typical separation between defects. It also proves useful to work with $(\gamma v)$. One can then change variables using
\be
\frac{d\gamma}{dt}=v\gamma^3\frac{dv}{dt}\,.
\ee
Noting that in the canonical model the curvature radius $R$ is an invariant quantity, which in a one-scale context is identified as $R_{inv}\equiv L_{inv}$, and transforming it to the physical one, we finally obtain
\be
\frac{d(\gamma v)}{dt}=\frac{k\gamma^{1+\frac{1}{4-n}}}{L_{ph}}-nH{\gamma v}
\ee
\be
(4-n)\frac{dL_{ph}}{dt}=(4-n)HL_{ph}+v(k+c)\gamma^{\frac{1}{4-n}}\,.
\ee
If we now look for attractor scaling solutions we get
\be
\epsilon^2_{ph}=\gamma^{\frac{2}{4-n}}\epsilon^2_{inv}
\ee
\be
(\gamma v)^2_{ph}=\gamma^2 v^2_{inv}\,,
\ee
which is trivially correct and consistent given the various definitions above. Finally one needs to confirm how the model parameters $c$ and $k$ are transformed as one switches between the physical and invariant approaches. Again, one can show \cite{Cabral} that
\be
c_{ph}=\gamma^{\frac{1}{4-n}}c_{inv}
\ee
\be
k_{ph}=\gamma^{\frac{1}{4-n}}k_{inv}\,;
\ee
in the former case this can be rigorously derived following an argument by Kibble \cite{Kibble2} while in the latter the relation is more phenomenological---which is reflection of the phenomenological nature of the parameter $k$ itself \cite{MartinsShellard2}.

With these relations between the physical and invariant model parameters, we can finally write
\be\label{physL}
(4-n)\frac{dL_{ph}}{dt}=(4-n)HL_{ph}+(c_{ph}+k_{ph})v\,
\ee
\be\label{physV}
\frac{dv}{dt}=(1-v^2)\left[\frac{k_{ph}}{L_{ph}}-nHv\right]\,,
\ee
or equivalently
\be
\frac{d(\gamma v)}{dt}=\frac{\gamma k_{ph}}{L_{ph}}-nH{\gamma v}\,,\label{physgv}
\ee
which are the evolution equations for the VOS model based on physical rather than invariant parameters.


\section{Radiation-to-matter transition}

In this section we will obtain an approximate analytic solution for the behavior of a defect network satisfying the generic evolution equations given by Eqs. \ref{physL}-\ref{physV} during the transition from radiation domination to matter domination. We start by noting that the exact solution for the behavior of the scale factor during this transition is \cite{Kolb}
\be\label{aRM}
\frac{a(\tau)}{a_{eq}}=\left(\frac{\tau}{\tau_*}\right)^2+2\left(\frac{\tau}{\tau_*}\right)\,,
\ee
where $\tau$ is conformal time, related to physical time $t$ via
\be\label{physconf}
dt=a\,d\tau\,,
\ee
and we have defined
\be
\tau_*=\frac{\tau_{eq}}{\sqrt{2}-1}\,,
\ee 
with $a_{eq}$ and $\tau_{eq}$ respectively denoting the scale factor and conformal time at the epoch of equal radiation and matter densities.

\begin{figure}
\begin{center}
\includegraphics[width=3.1in]{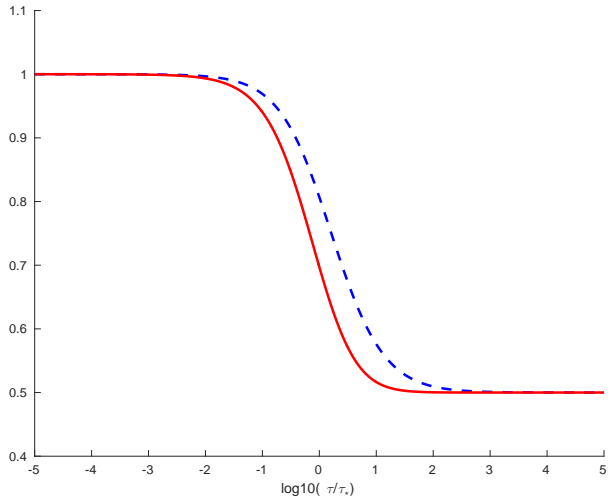}
\end{center}
\caption{\label{fig1}The behavior of the time-dependent factors on the right-hand side of Eq. \protect\ref{tdfL} (blue dashed line) and Eq. \protect\ref{tdfV} (solid red line).}
\end{figure}

Given the simplicity of this solution it is useful to rewrite the VOS dynamical equations in terms of conformal time,  which can be straightforwardly done using Eq. \ref{physconf}. Moreover, it is also convenient to work with the comoving version of the physical length scale $L_{ph}$, defining
\be
L_{ph}=a\,\xi_{ph}\,.
\ee
We therefore find
\be\label{cmvL}
(4-n)\frac{d\xi_{ph}}{d\tau}=(c_{ph}+k_{ph})v
\ee
\be\label{cmvV}
\frac{dv}{d\tau}=(1-v^2)\left[\frac{k_{ph}}{\xi_{ph}}-n{\cal H}v\right]\,,
\ee
where ${\cal H}=d\ln{a}/d\tau$ is the comoving Hubble parameter. From these one can similarly show that the attractor scaling solution is
\be
\xi_{ph}=\zeta \tau\,,\quad v=v_0=const.
\ee
with
\be\label{stdcomL}
\zeta^2=m^2_{ph}\frac{1-\lambda}{\lambda}\,,
\ee
where 
\be
m^2_{ph}=\gamma^{\frac{2}{4-n}}m^2
\ee
is the physical analog of the invariant quantity defined in Eq. \ref{defm2}, and
\be\label{stdcomV}
v_0^2=\frac{4-n}{n}\frac{1-\lambda}{\lambda}\frac{k_{ph}}{k_{ph}+c_{ph}}\,.
\ee
As expected these exactly match the previously obtained ones, Eqs. \ref{stdL}-\ref{stdV}: the solutions for the velocities are exactly the same, and the ones for the correlation length also match since for a power-law expansion ($a\propto t^\lambda$)
\be
\frac{L_{ph}}{t}=\epsilon_{ph}=\frac{1}{1-\lambda}\frac{\xi_{ph}}{\tau}=\frac{\zeta}{1-\lambda}\,.
\ee

A simple solution can now be found for the transition between radiation and matter domination. We expect $c_{ph}$ and $k_{ph}$ to either be constants or slowly varying parameters, and based on the results of both Goto-Nambu and field theory simulations of cosmic strings \cite{Abelian,Goto,BlancoPillado,Kunz}, as well as of field theory simulations of domain walls \cite{Rybak}, we further expect the defect velocities to change slowly during the transition---a point already anticipated by Kibble \cite{Kibble2}. This implies that the $dv/d\tau$ term in Eq. \ref{cmvV} should also be small, and therefore this equation yields
\be\label{vaprox}
v\xi_{ph}{\cal H}=\frac{k_{ph}}{n}\,.
\ee
This can now be substituted in Eq. \ref{cmvL}, leading to
\be
\xi_{ph}\frac{d\xi_{ph}}{d\tau}=m^2_{ph}\frac{1}{\cal H}\,.
\ee
Since for the radiation-to-matter transition there is an exact analytic expression for ${\cal H}(\tau)$, which can be trivially calculated from Eq. \ref{aRM}, we can now integrate this equation. The result is
\be\label{tdfL}
\zeta^2(y)=m^2_{ph}\left[\frac{1}{2}+\frac{y-\ln{(1+y)}}{y^2}  \right]
\ee
where for the sake of simplicity we have further defined a dimensionless conformal time $y=\tau/\tau_*$. Substituting this solution in Eq. \ref{vaprox} we find for the velocity
\be\label{tdfV}
v^2(y)=\frac{4-n}{n}\frac{k_{ph}}{k_{ph}+c_{ph}}\, \frac{y^2\left(2+y\right)^2}{4\left(1+y\right)^2\left[y+\frac{1}{2}y^2-\ln{\left(1+y\right)} \right] }\,.
\ee
It is straightforward to show that in the limits of small and large conformal times (corresponding to the deep radiation and matter eras, respectively) these expressions reduce to the ones given by Eqs. \ref{stdcomL}-\ref{stdcomV}.

Figure \ref{fig1} depicts the behavior of the time-dependent factors on the right-hand side of Eqs. \ref{tdfL}-\ref{tdfV}: both of them are unity in the deep radiation era and $1/2$ in the deep matter era. It is noteworthy that the transition takes about 5 orders of magnitude of conformal time, and that the transition occurs somewhat earlier for the velocities than it does for the correlation length (or equivalently the density). The reason for this is intuitively clear: as the expansion rate starts increasing relative to the $t^{1/2}$ radiation-era behavior the additional damping slows the defects down, and this affects their interaction rate (specifically, in the case of strings, the number of intercommutings and loop production events), which within the subsequent Hubble time will in turn be reflected in the network's density.


\section{Matter-to-acceleration transition}

We now turn to the transition between the matter-dominated era and the onset of acceleration. For the particular case in which the dark energy is due to a cosmological constant and with the further assumption of a flat universe ($\Omega_m+\Omega_\Lambda=1$) there is also an analytic solution for the behavior of the scale factor, though in this case as a function of physical time
\be
\left(\frac{a(t)}{a_0}\right)^{3/2}=\sqrt{\frac{\Omega_m}{\Omega_\Lambda}}\sinh{\left[\frac{3}{2}\sqrt{\Omega_\Lambda}H_0t\right]}\,,
\ee
where $a_0$ and $H_0$ are the present-day values of the scale factor and the Hubble parameter. We can, in principle, proceed under the same assumptions as in the previous section, and in this case we will obtain
\be
\xi\frac{d\xi}{dt}=m^2\frac{1}{a^2(t)H(t)}\,.
\ee
This equation could also be integrated, but the resulting integral would depend on hypergeometric functions and therefore it would not be particularly illuminating. Instead, we can change variables, using redshift instead of time. Thus we easily find
\be\label{evolacc}
\frac{d\ln{\xi}}{d\ln{(1+z)}}=-m^2\frac{(1+z)^2}{H(z)^2\xi^2}\,;
\ee
for the particular case of a flat $\Lambda$CDM model we have
\be
H(z)^2=H_0^2\left[\Omega_m(1+z)^3+\Omega_\Lambda \right]\,.
\ee
This can then be substituted in Eq. \ref{evolacc}, which can be integrated leading to
\be
H_0^2\xi^2(z)=m^2\, g(z)\,,
\ee
where $g(z)$ is a redshift-dependent function
\begin{eqnarray}\label{zdepL}
3\Omega_m^{2/3}\Omega_\Lambda^{1/3} & & g(z)= \nonumber \\
& & \log{\left[\frac{\left[\Omega_m^{1/3}(1+z)+\Omega_\Lambda^{1/3}\right]^2}{\Omega_m^{2/3}(1+z)^2-(\Omega_m\Omega_\Lambda)^{1/3}(1+z)+\Omega_\Lambda^{2/3}} \right]} \nonumber \\
& & -2\sqrt{3}\arctan{\frac{2\Omega_m^{1/3}(1+z)-\Omega_\Lambda^{1/3}}{\sqrt{3}\Omega_\Lambda^{1/3}}} + \sqrt{3}\pi \,,
\end{eqnarray}
and mathematically this expression assumes that both $\Omega_m$ and $\Omega_\Lambda$ are non-zero. The velocity $v(z)$ can again be obtained under the same approximation as before (cf. Eq. \ref{vaprox}), yielding
\be\label{zdepV}
v^2=\frac{4-n}{n}\frac{k}{k+c}\, \frac{(1+z)^2}{\left[\Omega_m(1+z)^3+\Omega_\Lambda  \right]g(z)}\,.
\ee
Note that despite keeping both $\Omega_m$ and $\Omega_\Lambda$ in the above in order to have slightly shorter expressions, we are assuming a flat universe, and therefore $\Omega_m+\Omega_\Lambda=1$. For non-flat universes the evolution equations of the VOS model will include curvature-dependent terms \cite{MartinsShellard2}.

\begin{figure}
\begin{center}
\includegraphics[width=3.1in]{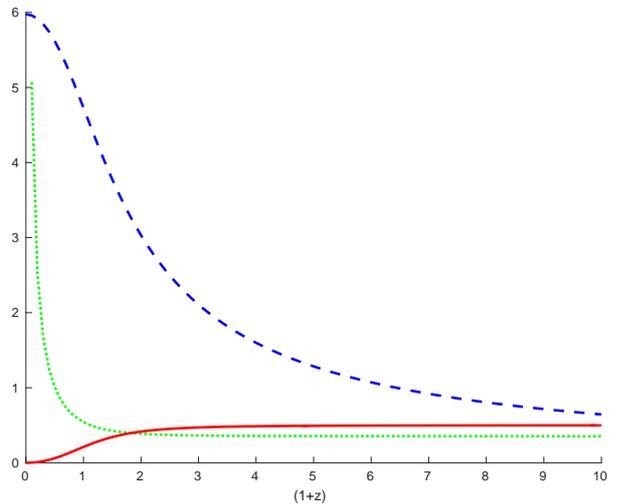}
\end{center}
\caption{\label{fig2}The behavior of the redshift-dependent function $g(z)$ defined in Eq. \protect\ref{zdepL} (blue dashed line), of the redshift-dependent part of the right-hand side of Eq. \protect\ref{zdepV} (solid red line), and of $\theta/(4m)$ (cf. Eq. \protect\ref{thetahere}, green dotted line).}
\end{figure}

For both of these equations one can show that in the limit $z\longrightarrow\infty$ one recovers the matter-era linear scaling solution, while in the opposite limit $(1+z)\longrightarrow 0$ (corresponding to the far future) one obtains the inflating solution where the network is frozen and conformally stretched, with $L\propto a$ and $v\propto a^{-1}$ \cite{MartinsShellard2}. It is worth pointing out that physically this is the correct way to obtain the matter-era and inflationary solutions.

Figure \ref{fig2} depicts the behavior of $g(z)$ defined in Eqs. \ref{zdepL} and of the redshift-dependent part of the right-hand side of Eq. \ref{zdepV}. Note that for large redshifts $g(z)\propto a\propto (1+z)^{-1}$, while it tends to a constant in the far future. For this reason it is also useful to consider the dimensionless parameter
\be\label{deftheta}
\theta=HL\,,
\ee
which has the opposite behavior: it is a constant in the deep matter era and it grows when the universe accelerates. In terms of our previously defined parameters, this can be written as
\be\label{thetahere}
\theta=\frac{m}{1+z}\frac{H(z)}{H_0}\sqrt{g(z)}\,.
\ee
Note that $g(z)$ is independent of the defect dimension but $v$ and $\theta$ will depend on it. Figure \ref{fig2} also depicts the behavior of $\theta/(4m)$, the factor of $1/4$ being chosen simply in order to make it more visible within the range of variation of the other parameters.

\begin{figure*}
\begin{center}
\includegraphics[width=3.5in]{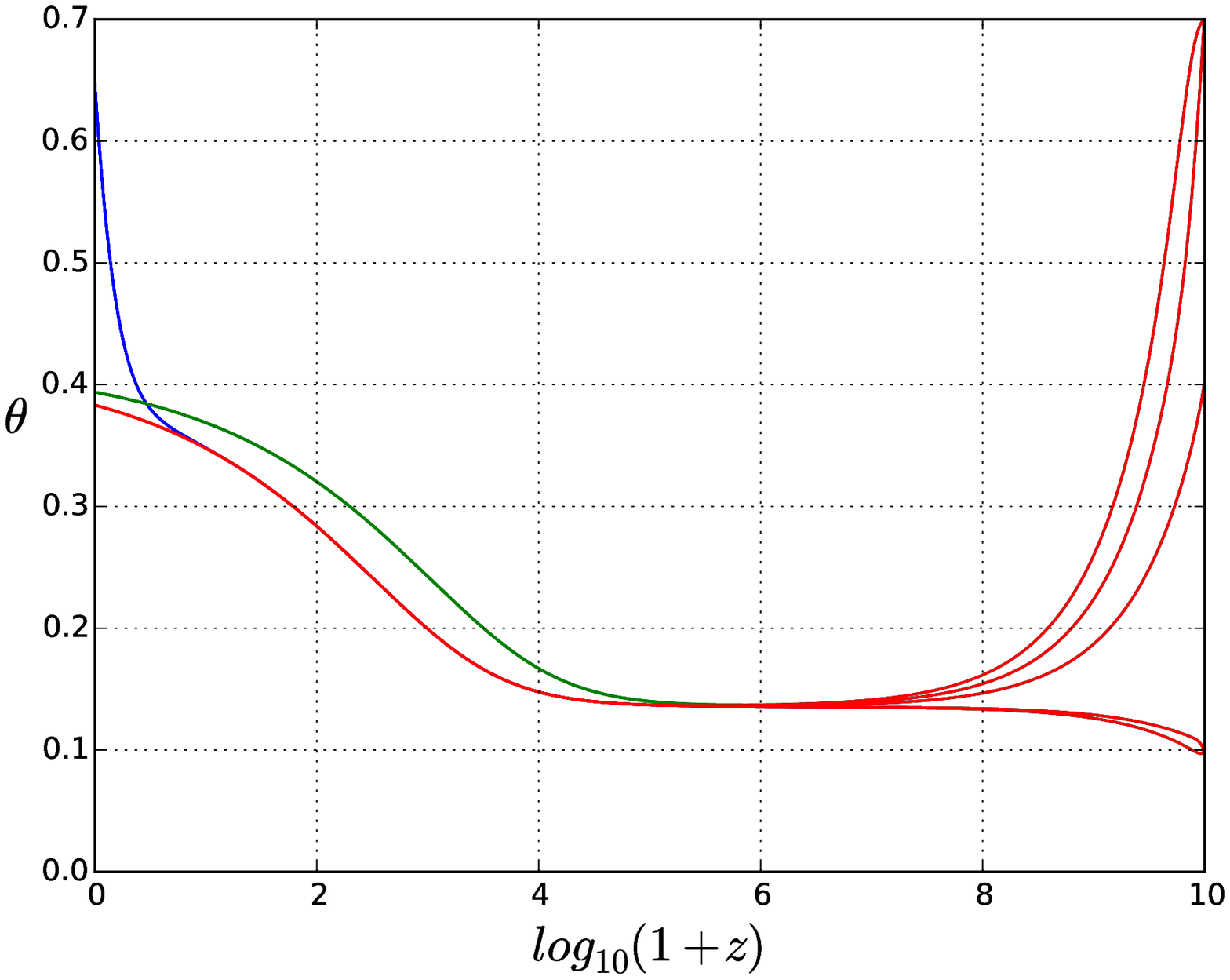}
\includegraphics[width=3.5in]{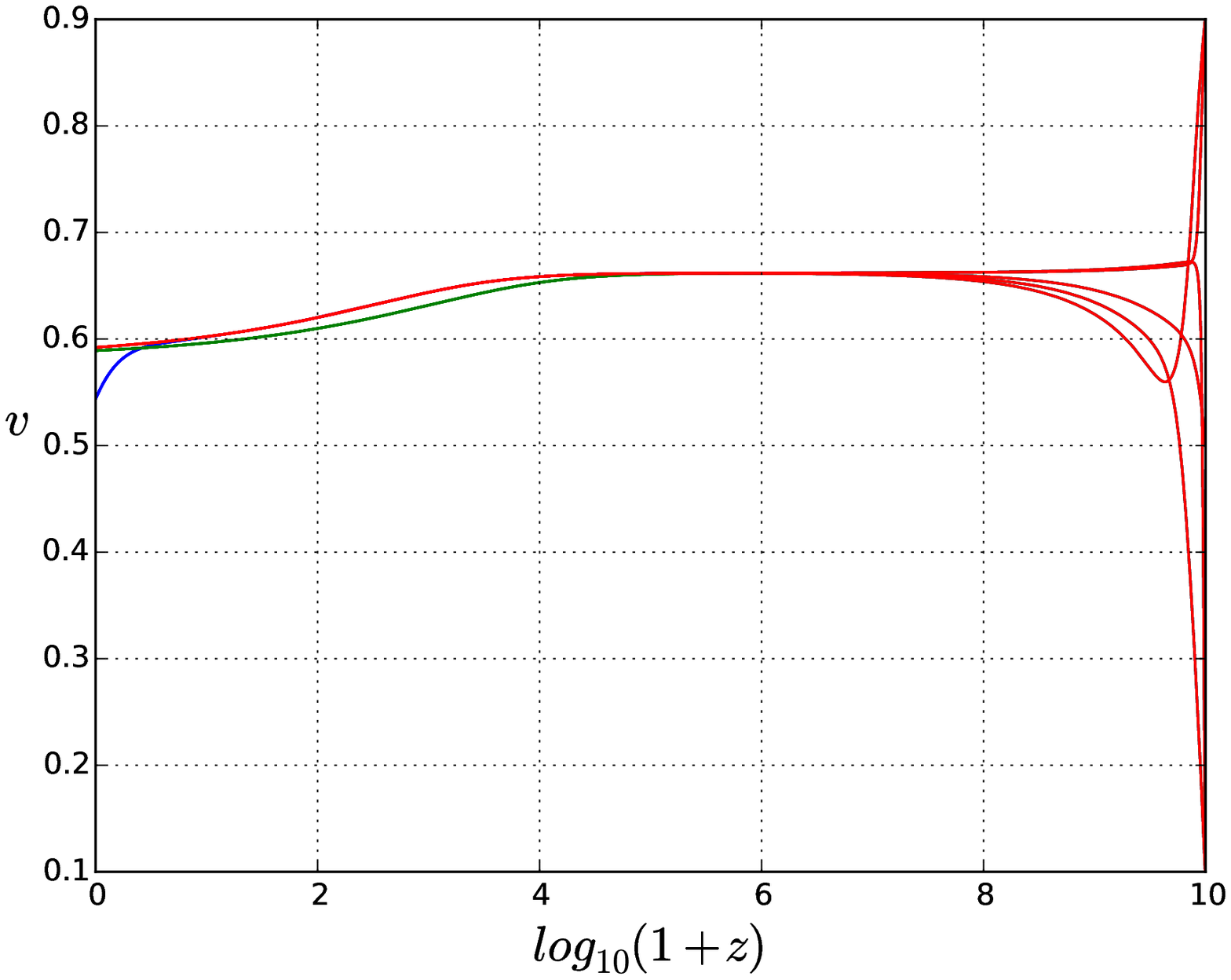}
\includegraphics[width=3.5in]{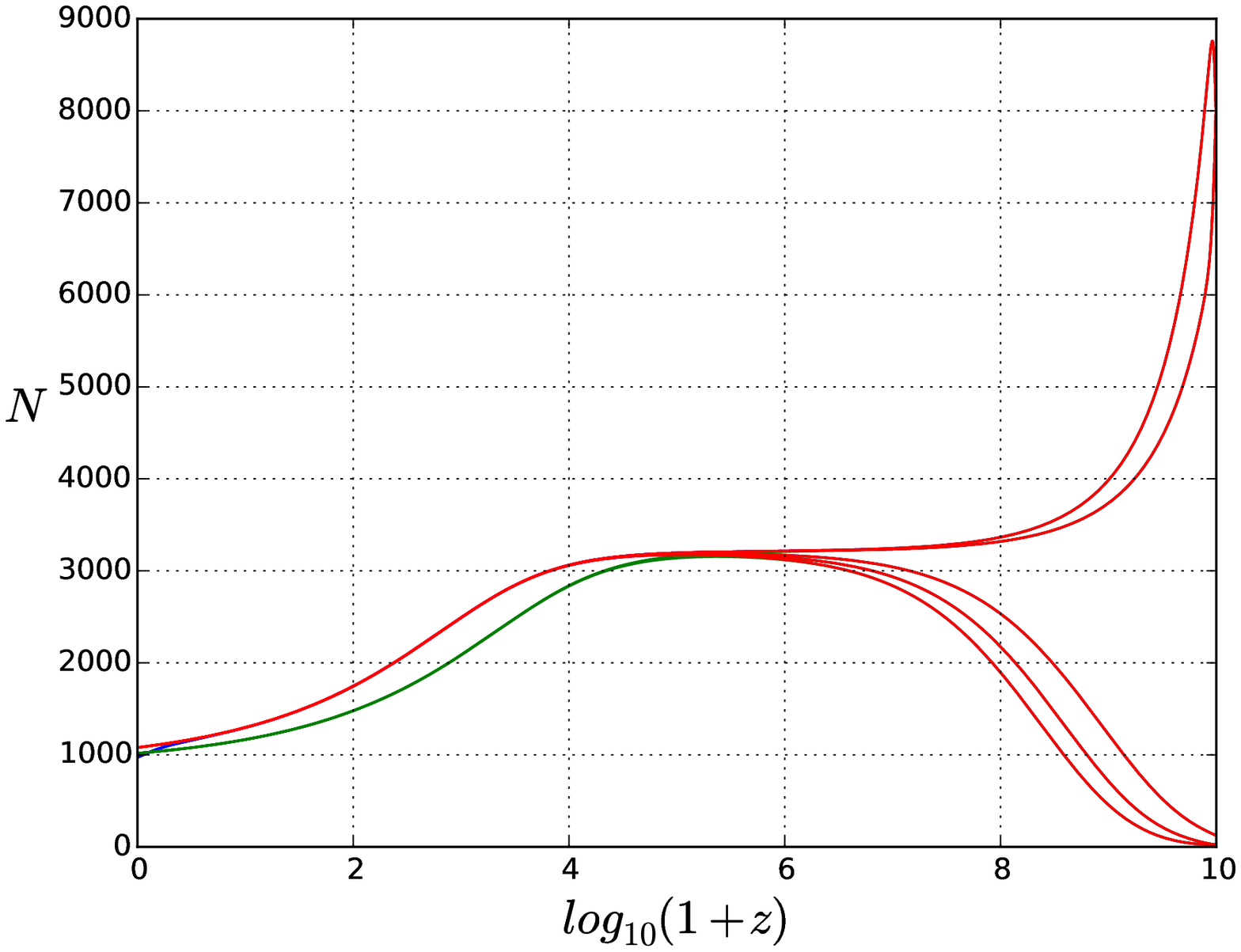}
\end{center}
\caption{\label{fig3}Comparing the evolution of the VOS model parameters $\theta(z)$ and $v(z)$ and the number of string segments $N$ in models containing the same amount of radiation, but different amounts of matter and dark energy: the green line shows a flat matter-only universe, the red ones show an open universe, and the blue one shows a flat $\Lambda$CDM universe. All models contain the same amount of radiation $\Omega_r\sim9\times10^{-5}$, and the parameters in the $\Lambda$CDM case are in agreement with the recent Planck satellite data \protect\cite{Ade}; see the main text for specific parameter values. Each model is integrated starting with various different initial conditions, representative of high and low network densities and velocities, showing that the initial conditions are erased well before the radiation-to-matter transition. Note that the green and red curves diverge due to the radiation-to-matter transition, while the blue and red curves diverge due to the onset of the acceleration of the universe.}
\end{figure*}

Since one of our goals is to study how the network properties depend on the underlying dark energy model, a more general numerical approach will be used in what follows. For a fiducial class of dark energy models we will use the standard Chevallier-Polarski-Linder (CPL) parametrization
\begin{equation}\label{eq:omega}
w(z)=w_0+w_a\frac{z}{1+z},
\end{equation}
corresponding to the Friedmann equation
\begin{eqnarray} 
\label{eq:friedmann_red}
\frac{H^2(z)}{H_0^2} &=& \Omega_r(1+z)^4+\Omega_m (1+z)^3 \nonumber \\
 &+& (1-\Omega_r-\Omega_m)(1+z)^{3(1+w_0+ w_a)} e^{-\frac{3w_a z}{1+z}}\,,
\end{eqnarray} 
where $\Omega_r\sim9\times10^{-5}$ is the present-day value of the fraction of the critical density in radiation and we are still assuming a flat universe (consistently with the latest observational data). We will simultaneously integrate the length scale and velocity equations, written as a function of redshift; in full generality we have
\be
- \frac{d\ln{L}}{d\ln{(1+z)}}=1+\frac{nv^2}{4-n}+\frac{cv}{(4-n)HL}
\ee
\be
-\frac{dv}{d\ln{(1+z)}}=(1-v^2)\left(\frac{k(v)}{HL}-nv\right)\,.
\ee
In our particular case we are interested in cosmic strings (corresponding to $n=2$) and for convenience of representing the numerical results will replace the length scale $L$ by the dimensionless parameter $\theta$ defined in Eq. \ref{deftheta}. Therefore the system of equations to be numerically integrated is
\be
\frac{d\ln{\theta}}{d\ln{(1+z)}}= \frac{d\ln{H(z)}}{d\ln{(1+z)}}-(1+v^2)-\frac{cv}{2\theta}
\ee
\be
\frac{dv}{d\ln{(1+z)}}=(1-v^2)\left(2v-\frac{k(v)}{\theta}\right)\,,
\ee
with $H(z)$ given by the Friedmann equation (cf. Eq. \ref{eq:friedmann_red}) and the momentum parameter given by \cite{MartinsShellard2}
\be\label{defnk}
k(v)=\frac{2\sqrt{2}}{\pi}(1-v^2)(1+2\sqrt{2}v^3)\frac{1-8v^6}{1+8v^6}\,.
\ee

\begin{figure*}
\begin{center}
\includegraphics[width=3.3in]{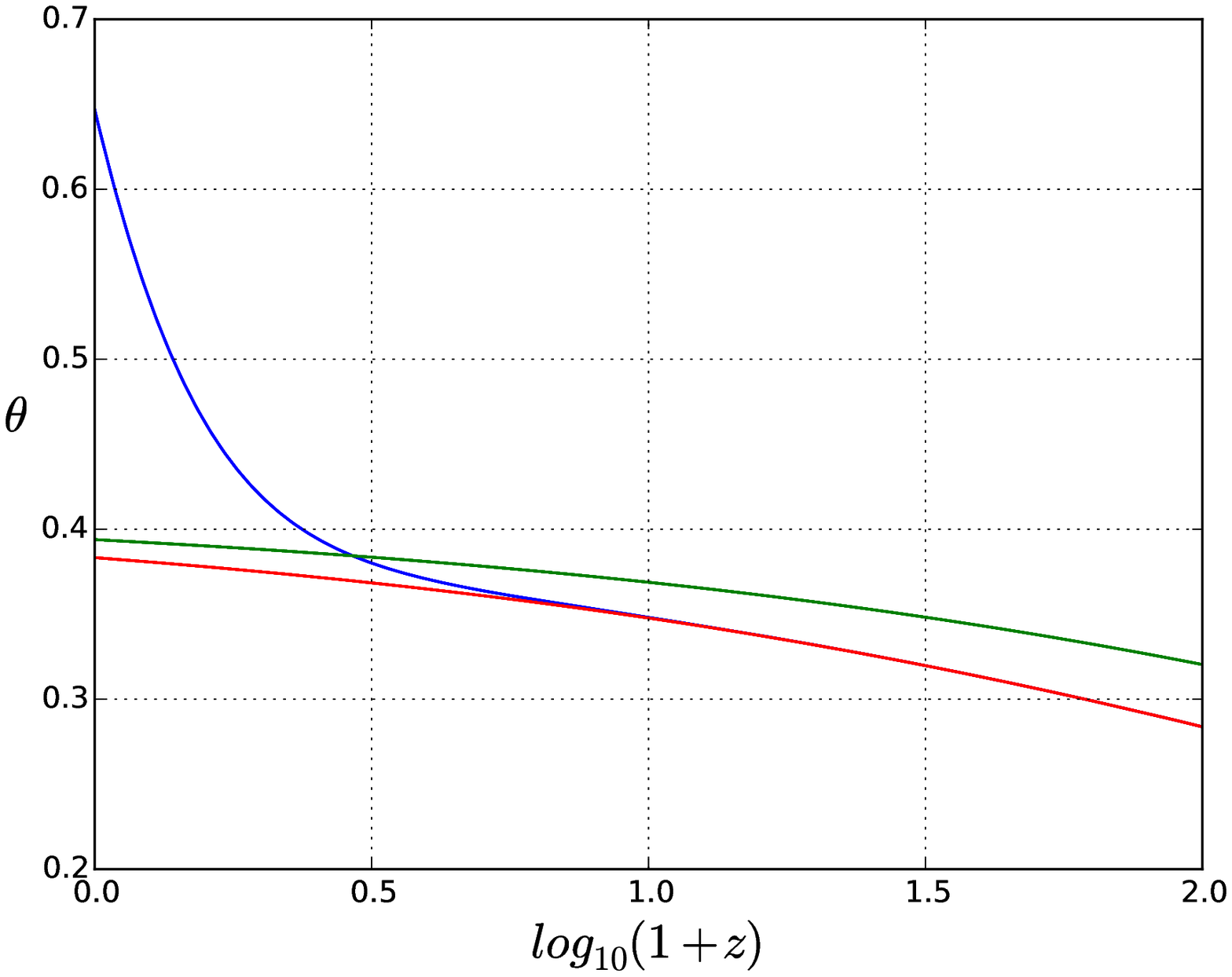}
\includegraphics[width=3.3in]{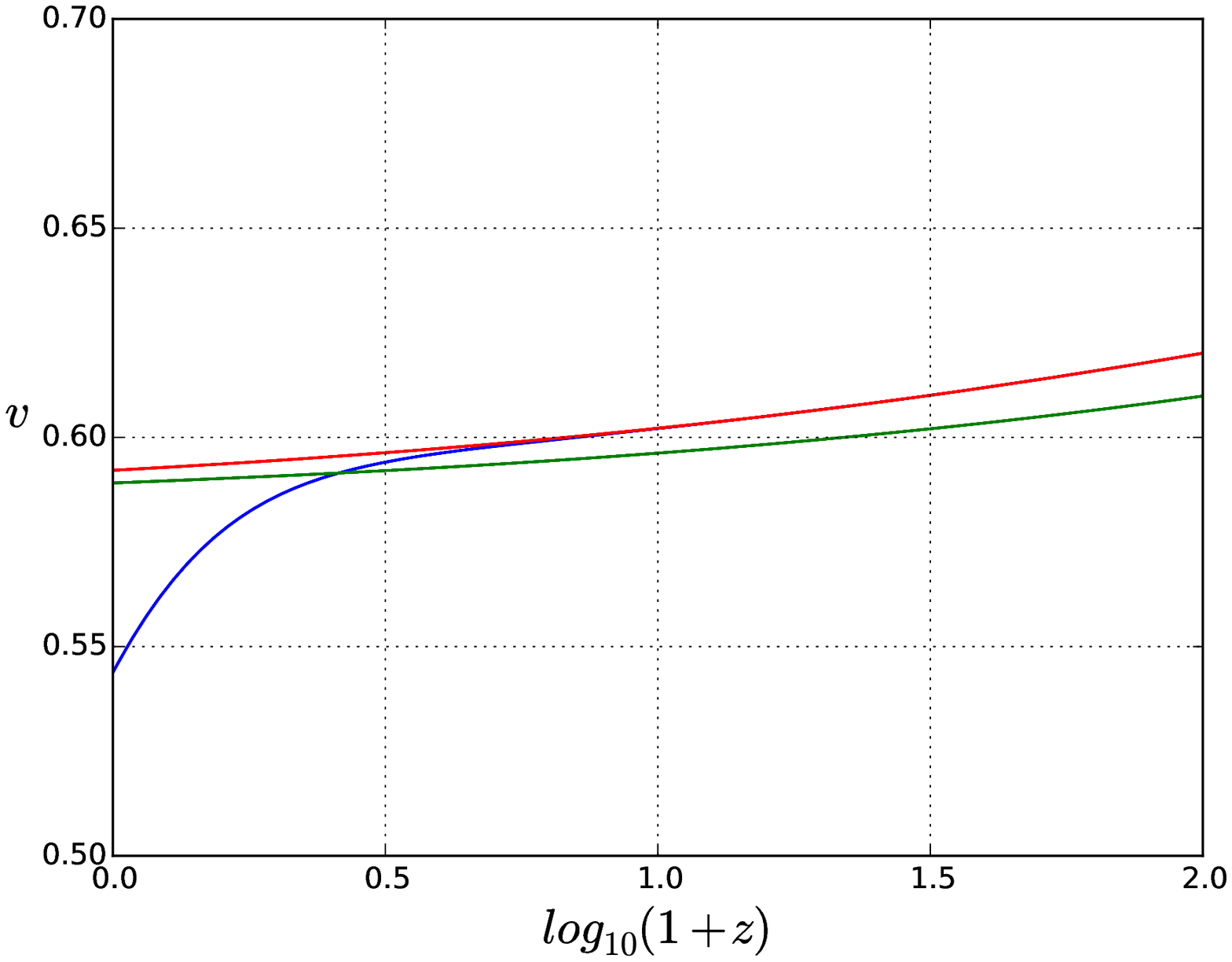}
\includegraphics[width=3.3in]{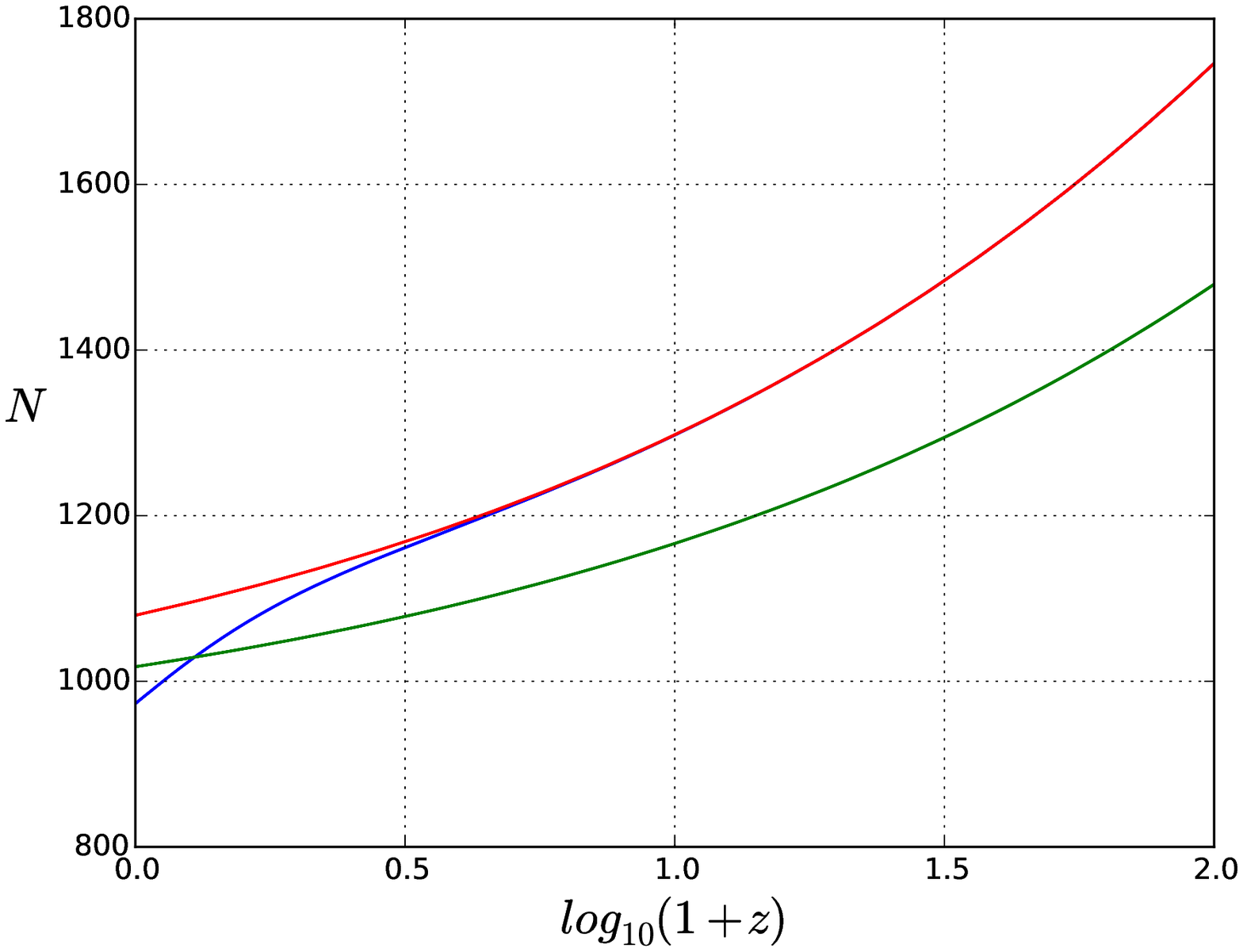}
\end{center}
\caption{\label{fig4}Same as Fig. \protect\ref{fig3}, highlighting the behavior close to the present day.}
\end{figure*}
\begin{figure*}
\begin{center}
\includegraphics[width=3.3in]{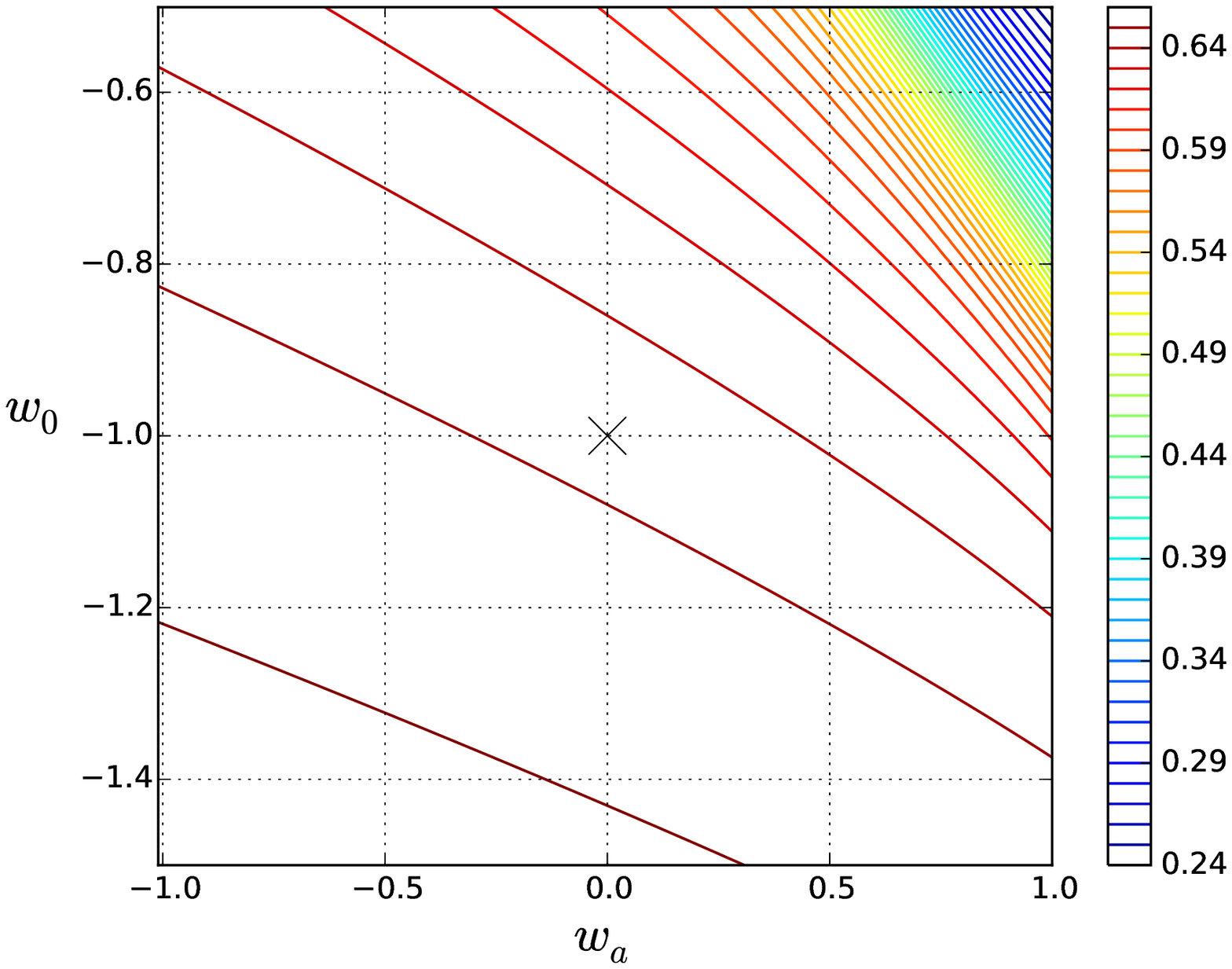}
\includegraphics[width=3.3in]{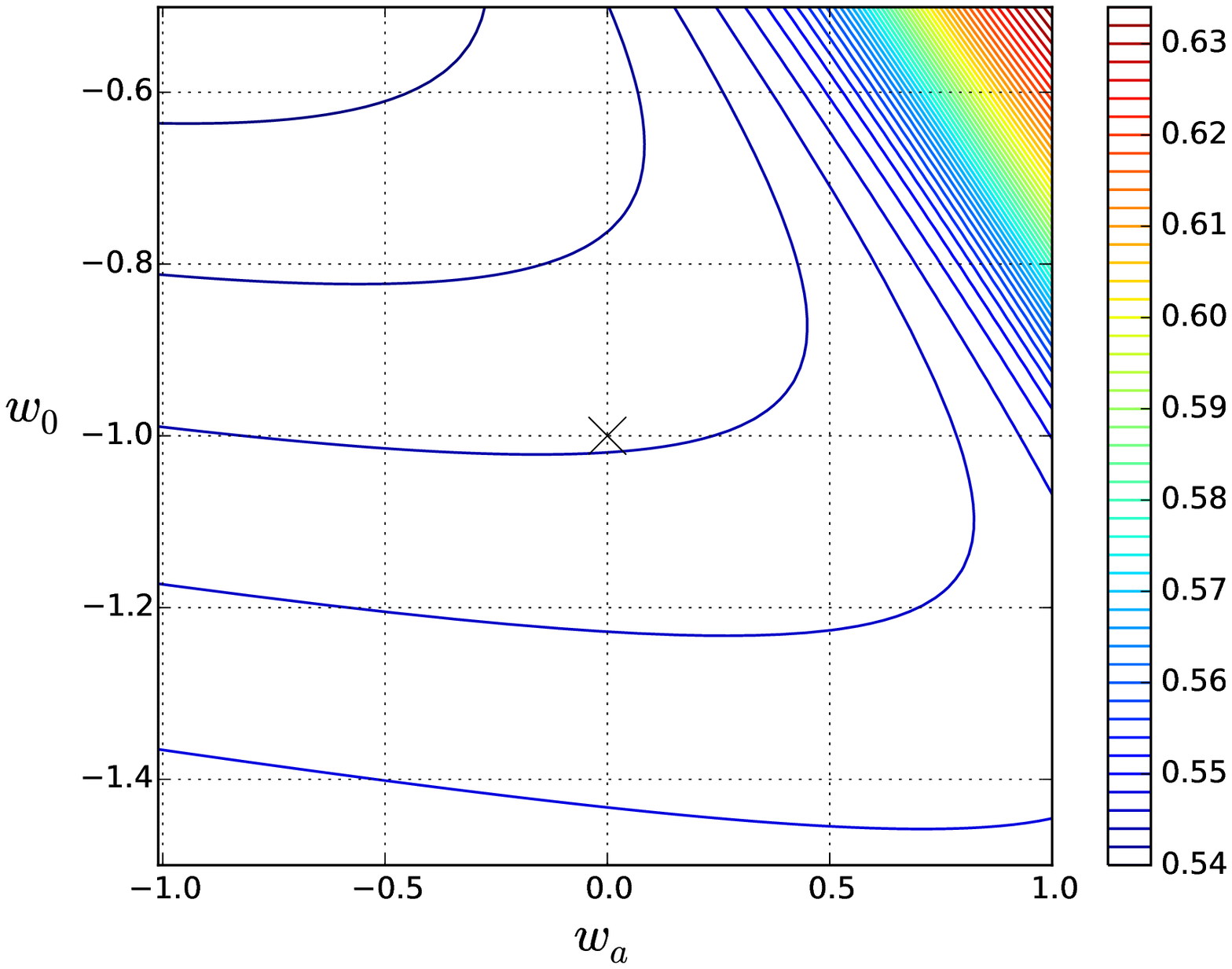}
\includegraphics[width=3.3in]{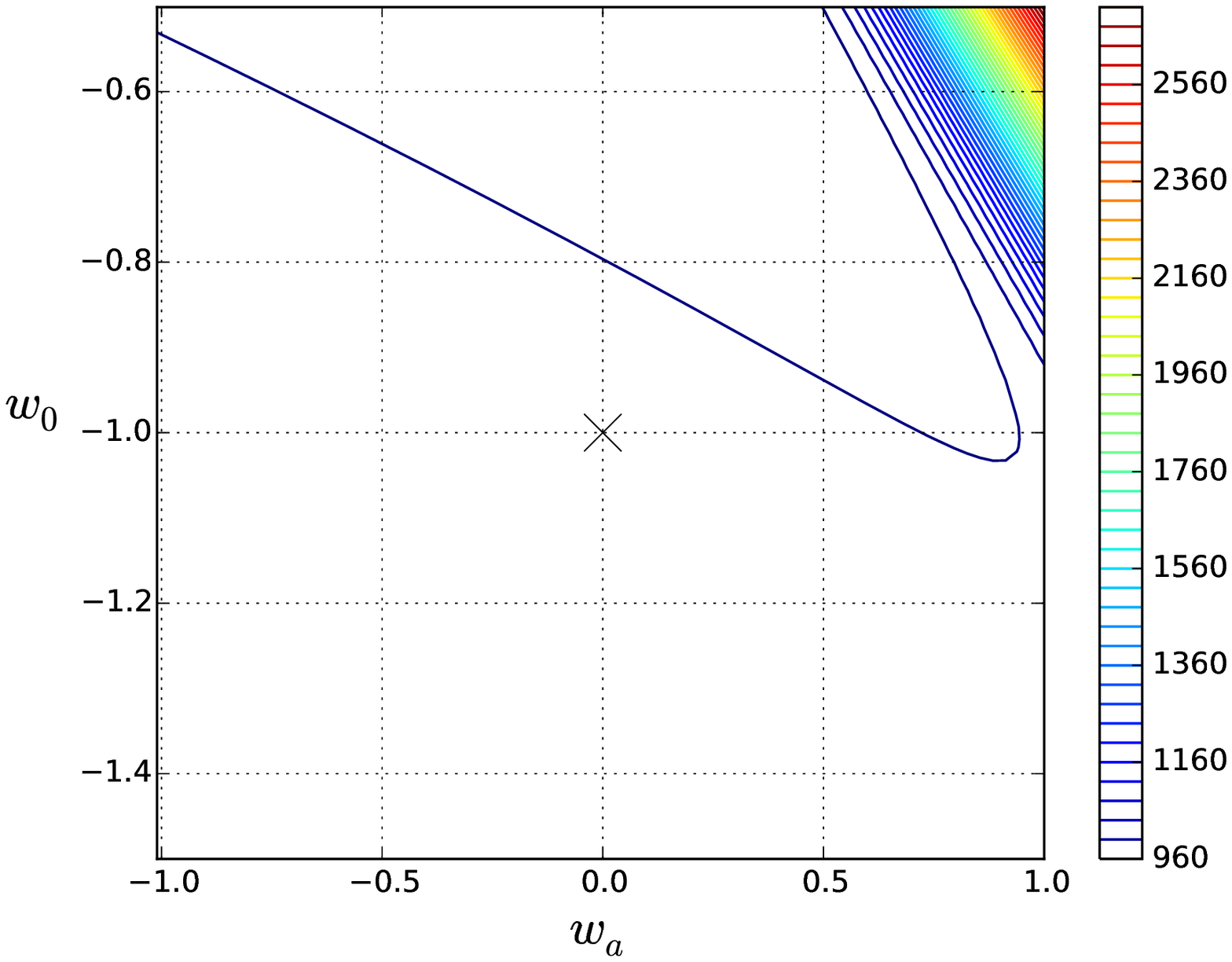}
\end{center}
\caption{\label{fig5}Dependence of the present-day values of the parameters $\theta$, $v$ and $N$ (shown on the top left, top right and bottom panels, respectively) on the dark energy equation of state, described by the CPL parametrization---cf. Eq. \protect\ref{eq:omega}. The parameter values associated with each contour are shown on each color bar. The cross highlights the $\Lambda$CDM case, corresponding to $w_0=-1$ and $w_a=0$.}
\end{figure*}

Figure \ref{fig3} compares the evolution of three cosmological models with the same amount of radiation ($\Omega_r=9.2\times10^{-5}$) but different amounts of matter and dark energy (in the form of a cosmological constant); for the $\Lambda$CDM case, the chosen values of the cosmological parameters are in agreement with the recent Planck satellite data \cite{Ade}. Specifically we have
\begin{itemize}
\item Blue line: flat $\Lambda$CDM model, with $\Omega_m=0.3089$ and $\Omega_\phi=1-\Omega_m-\Omega_r$
\item Green line: flat model with $\Omega_\phi=0$ and $\Omega_m=1-\Omega_r$.
\item Red lines: open model with $\Omega_\phi=0$ and $\Omega_m=0.3089$
\end{itemize}
We do the integrations with different initial conditions, representative of networks with high and low densities and velocities, specifically the four permutations of $\theta=0.1, 0.7$ and $v=0.1, 0.9$, as well as an intermediate case with $\theta=0.4$ and $v=0.5$. Physically, different values of the initial density (hence, correlation length) and velocity would depend, for example, on the order of the string-forming phase transition \cite{Vilenkin}. For numerical convenience we start integrating at redshift $z=10^{10}$; although this is much closer to the present day than, say, the epoch of formation of a grand unified theory-scale network, the figure shows that the initial conditions are erased before the radiation-to-matter transition starts, so this is fine for our purposes in the present work.

Comparing the three models one finds that the differences between the behaviors of $\theta$ and $v$ are small, while for $N$, which involves a comparison with the horizon and therefore depends on the expansion history, they are somewhat larger. Note that at early time (large redshifts) the green and red curves overlap: they become different around the radiation-to-matter transition (which occurs at different redshifts in the two models, since both have the same amount of radiation but different amounts of matter). Similarly, the red and blue curves (which have the same amounts of radiation and matter, but different amounts of dark energy become different with the onset of acceleration. Figure \ref{fig4} depicts a zoomed version of the low-redshift behavior. For $\Lambda$CDM the number of strings in the present-day observable universe is expected to be
\be
N_0=973.0\pm0.6\,,
\ee
where the error bar has been numerically estimated from the one-sigma uncertainties coming from the relevant cosmological parameters \cite{Ade} and also takes into account the network's initial conditions (although the latter are clearly subdominant).

That said, we should also point out that in integrating these equations we have kept the energy loss parameter $c$ fixed at the value $c=0.23$ obtained from calibration against numerical simulations \cite{Abelian,Goto}, while for $k$ we have used the phenomenological formula given by Eq. \ref{defnk}. Naturally there are also uncertainties associated with these parameters (estimated to be at the $10\%$ to $20\%$ level, though they are difficult to quantify in detail), and propagating these uncertainties would increase the above uncertainty in $N$. In any case, the clear bottom line of this analysis is that when calculating detailed observational consequences of cosmic strings one should keep in mind that a realistic network will never be exactly scaling in the observationally relevant redshift range. 

Finally we study how the present-day values of the observable parameters describing the string network depend on the dark energy equation of state, described by the CPL parametrization in Eq. \ref{eq:omega}. In doing so only the CPL parameters $w_0$ and $w_a$ are varied, with the present-day values of the matter and dark energy densities kept fixed at the best-fit values of the Planck satellite data \cite{Ade}. For these integrations we use initial conditions $\theta=0.4$ and $v=0.5$, but as our previous analysis demonstrates this specific choice does not affect the results since the initial conditions are erased by the evolution well before the present day.

Figure \ref{fig5} shows the results of this analysis. A changing equation of state will affect both the speed of the matter-to-acceleration transition and the redshift of its onset, and this is reflected in the dynamics of the string network. Naturally, for large values of $w_0$ and $w_a$ the universe would never start accelerating and the string network would still have a behavior similar to that of the matter-era scaling solution (but such large values are observationally excluded). On the other hand, for the region of the $w_0$--$w_a$ parameter space in the neighborhood of the cosmological constant ($w_0=-1$, $w_a=0$) the observable string parameters are relatively insensitive to the equation of state. Specifically for $\Lambda$CDM we find to four significant digits
\be
\theta_0 = 0.6468
\ee
\be
v_0 = 0.5438
\ee
\be
N_0 = 973.0\,,
\ee
in full agreement with our previous analysis---as can be seen by inspecting the blue lines in Fig. \ref{fig4}. So in this sense, assuming a background $\Lambda$CDM cosmology when exploring cosmological constraints on defect networks seems reasonably justified.

\section{Conclusions}\label{sec:conc}

It is well known that in idealized conditions, such as universes with a single component and a scale factor growing as a power law $a\propto t^\lambda$, networks of cosmic strings and several other topological defects evolve toward asymptotic scaling solutions where the average network velocity is a constant and the characteristic length scale grows as $L\propto t$. This is well established both through analytic calculations and detailed numerical simulations---see Ref. \cite{VOSbook} for a recent review. However, realistic cosmological models contain radiation, matter and dark energy. In those cases, one expects that string network will never be exactly scaling. In this work we have used the canonical VOS model for the evolution of defect networks to better quantify how these networks evolve.

Specifically, making use of an approximation originally suggested by Kibble in the context of friction-dominated universes \cite{Kibble2} we have obtained new analytic solutions of the VOS model for the behavior of the network during the radiation-to-matter and matter-to-acceleration transitions. In the latter case we have assumed the canonical $\Lambda$CDM model, but our subsequent numerical exploration of a wider range of dark energy models has confirmed that given the current constraints on the dark energy equation of state this is a reasonable approximation. A comparison of our analytic solutions to high-resolution numerical simulations is beyond our present scope, but will be performed in subsequent work.

Our results will be important for accurate studies of the cosmological implications of cosmic strings and topological defects as a whole. Future high-resolution cosmic microwave background experiments will certainly constrain---and may eventually detect---the predicted effects from these networks \cite{PRISM,Stage4}. Similarly, more comprehensive studies of energy loss mechanisms (most notably, for cosmic strings, loop production) and their gravitational wave emission will lead to quantitative predictions for the stochastic background and rates of specific transient defect-related events, as well as their specific fingerprints, that could be detected by the current facilities (such as LIGO or Virgo) or future ones (such as LISA or DECIGO) \cite{Siemens,Binetruy}. These issues will be addressed in future work.

\begin{acknowledgments}
This work was done in the context of project PTDC/FIS/111725/2009 (FCT, Portugal). CJM is also supported by an FCT Research Professorship, contract reference IF/00064/2012, funded by FCT/MCTES (Portugal) and POPH/FSE (EC).
\end{acknowledgments}

\bibliography{strings}
\end{document}